\documentclass[twocolumn,showpacs,preprintnumbers,amsmath,amssymb]{revtex4}

\usepackage{graphicx}
\usepackage{dcolumn}
\usepackage{bm}

\begin{document}
\title{Compressibility of a two-dimensional hole gas in tilted magnetic field}
\author{Maryam Rahimi, M.~R. Sakr, and S.~V. Kravchenko}
\affiliation{Physics Department, Northeastern University, Boston, Massachusetts 02115}
\author{S.~C. Dultz and H.~W. Jiang}
\affiliation{Department of Physics and Astronomy, University of California at Los Angeles, Los Angeles, California 90095}
\date{\today}
\begin{abstract}
We have measured compressibility of a two-dimensional hole gas in p-GaAs/AlGaAs heterostructure, grown on a (100) surface, in the presence of a tilted magnetic field.  It turns out that the parallel component of magnetic field affects neither the spin splitting nor the density of states.  We conclude that: (a)~$g$-factor in the parallel magnetic field is nearly zero in this system; and (b)~the level of the disorder potential is not sensitive to the parallel component of the magnetic field.
\end{abstract}

\maketitle

Recently, the compressibility of two-dimensional (2D) hole systems has been studied near the zero magnetic field ($B=0$) metal-insulator transition in p-GaAs/AlGaAs heterostructures \cite{dultz00,ilani00,ilani01}.  It has been shown that in the metallic regime the hole compressibility is {\it negative}, but it changes sign when the hole density is decreased below a critical value corresponding to the transition to the insulating state.  It is known that in some 2D systems (electrons in silicon, holes in GaAs), the application of a parallel magnetic field, $B_\parallel$, causes the resistivity to grow by orders of magnitude and eventually leads to a complete destruction of the metallic state (for a review and references, see Ref.\cite{abrahams01}).  On the other hand, in the 2D hole system in SiGe, the effect of $B_\parallel$ is negligible since the spins cannot be aligned by the parallel magnetic field \cite{senz99, coleridge02}.  The mechanism behind the giant positive magnetoresistance is therefore generally believed to be tied to the alignment of spins of the free carriers, although other mechanisms have also been proposed \cite{altshuler99,meir99,dassarma00}.  Compressibility measurements may shed light on the mechanism of the destruction of the metallic state by $B_\parallel$ since they can distinguish between different mechanisms, {\it e.g.}, between field-induced spin alignment, field-induced distortion of the Fermi surface \cite{dassarma00}, or increase in the disorder potential \cite{altshuler99}.  To the best of our knowledge, the influence of the parallel magnetic field on compressibility has never been studied.

In this paper, we report measurements of the compressibility of 2D holes in a p-GaAs/AlGaAs heterostructure grown on a (100) surface.  Tilting the magnetic field allowed us to study the influence of both perpendicular, $B_\perp$, and parallel components of the field on the compressibility.  Surprisingly, no noticeable effect of $B_\parallel$ was found: both the compressibility and the strength of the disorder potential in the system remained nearly independent of the parallel component of the magnetic field.  We attribute the absence of the effect to the strong anisotropy of the $g$-factor which has recently been predicted and observed in a p-GaAs/AlGaAs heterostructure \cite{winkler00}.  According to Ref.\cite{winkler00}, the $g$-factor is nearly zero in the (100) direction; therefore, the application of a parallel magnetic field has a negligible effect on spins.  To test this, we measured the influence of the parallel magnetic field on the Zeeman splitting and determined that the $g$-factor does not exceed $\sim10^{-2}$.

The hole compressibility, $\kappa$, is proportional to the quantum capacitance, $C_q$: $\kappa=c_q/(p_se)^2=p_s^{-2}\,\frac{dp_s}{d\mu}$ (here $p_s$ is the density of 2D holes, $c_q\equiv C_q/A$ is the quantum capacitance per area, $A$; $\mu$ is the chemical potential, and $\frac{dp_s}{d\mu}$ is the thermodynamic density of states).  We studied the compressibility by measuring the electric field penetration through the 2D hole system, a method invented by Eisenstein {\it et al.} and used on double-layer electron system in GaAs \cite{eisenstein92} and later modified by Dultz and Jiang for use in heterostructures with a single layer of carriers \cite{dultz00}.  The penetrating field is related to the screening ability of the carriers and is inversely proportional to $\kappa$.  For example, infinite compressibility means that the electric field is completely screened by the hole system and is unable to penetrate through a 2D layer, finite positive compressibility corresponds to a partial penetration, and zero compressibility corresponds to a full penetration of the electric field.  Strong interaction between particles leads to a {\em negative} compressibility of a correlated 2D system \cite{bello81,efros88}: the direction of the penetrating field is opposite to the direction of the incident field.

We used a p-type MBE grown GaAs/Al$_{x}$Ga$_{1-x}$As single heterostructure with the maximum mobility of about $120\times10^3~\text{~cm}^2/\text{Vs}$ at a hole density of $p_s=2.60~\times~10^{11}\text{~cm}^{-2}$.  The device for compressibility measurements was fabricated by sandwiching the 2D hole gas between two metallic gates.  We applied a low-frequency ($f=$~0.2 to 5~Hz) excitation voltage $V_{ac}$ (typically 10~mV) to the bottom gate keeping the 2D hole system grounded in order to screen the electric field.  A DC voltage $V_g$ was superimposed to vary the hole density.  The current induced between the top gate and ground, proportional to the penetrating electric field, was measured by a lock-in amplifier.  Both the conductance, $\sigma$, and the quantum capacitance, $C_{q}$, of the 2D system can be extracted independently by measuring in-phase ($I_{x}$) and $90^o$ out-of-phase ($I_{y}$) current components, respectively.  In the low-frequency limit, $I_{x}$ is proportional to $\omega^2/\sigma$ and $I_{y}$ is proportional to $\omega/C_q$.  Special attention was paid to ensure that $\sigma$ and $C_q$ were frequency-independent.  Measurements were made in an Oxford dilution refrigerator equipped with a rotator.  Details of the sample preparation and experimental setup can be found in Ref.\cite{dultz00}.

Typical traces of both current components in zero magnetic field are shown in Fig.~1(a) as a function of the gate voltage; the latter determines the hole density via $p_s=(22.3-6.54{\text V_g})\cdot10^{10}$~cm$^{-2}$ (note that an increase in the gate voltage corresponds to a decrease in hole density).  $I_y$, proportional to the inverse compressibility, is negative at $V_g<2.75$~V (higher hole densities) and becomes ``more negative'' as $V_g$ is increased.  At a certain gate voltage $V_g=V_c$, however, $I_y$ sharply changes behavior and starts to rapidly increase, being accompanied by divergence of $I_x$ (proportional to the inverse conductivity).  This is in agreement with previous results obtained in electron \cite{eisenstein92} and hole \cite{dultz00,ilani00,ilani01} systems.  According to Refs.\cite{dultz00,ilani00,ilani01}, the abrupt change in the behavior of compressibility corresponds to the critical hole density, $p_c$, for the metal-insulator transition in this system.

When a perpendicular magnetic field is applied (Fig.~1(b)) peaks appear in both $I_y$ and $I_x$ traces corresponding to compressibility/conductivity minima at integer filling factors, $\nu=p_sch/eB_\perp=$~1, 2, 3...  Even filling factors correspond to the ``cyclotron gaps'' between neighboring Landau levels; odd ones correspond to the ``spin gaps'' between spin-up and spin-down levels within the same Landau level.  Note that at this relatively low magnetic field of 3~tesla the compressibility remains negative even at integer filling factors.
\begin{figure}
\scalebox{0.35}{\includegraphics{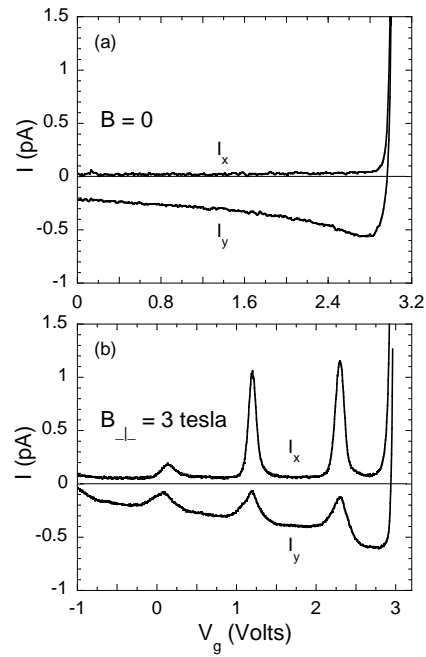}}\vspace{-.8cm}
\caption{\label{fig.1.} Typical traces of $I_{x}$ and $I_{y}$ {\it vs.}\ the gate voltage in zero magnetic field~(a) and in a perpendicular field $B_\perp=3$~tesla~(b).  $T=0.2$~K, $V_{ac}=20$~mV, $f=5$~Hz.}
\end{figure}

By integrating $I_y\propto d\mu/dp_s$ over $p_s$, we obtain the chemical potential versus hole density.  These dependences are shown in Fig.~2 (shifted vertically for clarity).  The chemical potential decreases for increasing $p_s$ in a similar way, both in zero magnetic field and in $B_\perp=3$~tesla.  In the latter case, weak oscillations corresponding to the Landau quantization can be seen in the $\mu(p_s)$ curve.

\begin{figure}\vspace{-28.5mm}
\scalebox{0.38}{\includegraphics{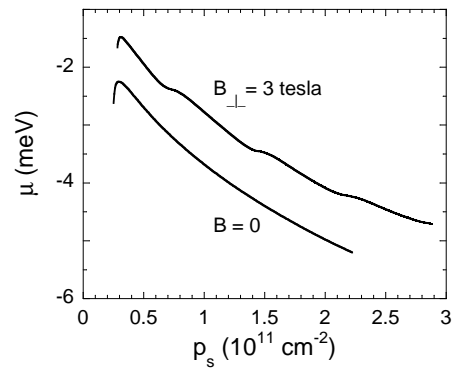}}\vspace{-2.5cm}
\caption{\label{fig.2.} Solid curves: Experimental results for the chemical potential {\it vs.}\ the hole density for $B=0$ (lower curve) and $B_\perp=3$~tesla (upper curve). 
The curves are vertically shifted for clarity.  $T=0.2$~K.}
\end{figure}

Application of a {\em parallel} magnetic field has negligible effect on both current components, as can be seen from Fig.~3.  While the effect of $B_\parallel$ on the compressibility has never been studied before, the absence of a field dependence of $I_x$ is unexpected: it is known that in ``conventional'' p-GaAs/AlGaAs heterostructures grown at a (311) surface the application of a parallel magnetic field causes giant magnetoresistance \cite{simmons98,yoon00} by destroying the metallic conductivity.  Since this giant magnetoresistance in generally attributed to spin effects, the observed difference may be attributed to anisotropy of the $g$-factor in this 2D system \cite{winkler00}.  Below, we verify this conjecture.

\begin{figure}\vspace{-27mm}
\begin{center}
\scalebox{0.38}{\includegraphics{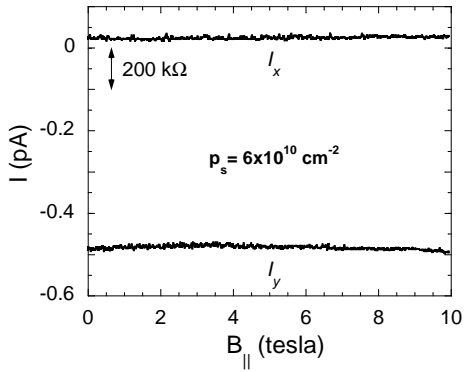}}\vspace{-2.8cm}
\end{center}
\caption{\label{fig.3.} $I_y$ and $I_x$ {\it vs.}\ parallel magnetic field for a fixed hole density of $p_s=6~\times~10^{10}\text{~cm}^{-2}$.  $T=0.2$~K, $B_\perp=0$, $V_{ac}=20$~mV, $f=5$~Hz.}
\end{figure}

In Fig.~4, we demonstrate that the compressibility, measured in the presence of a fixed $B_\perp$, is insensitive to the parallel component of the magnetic field: traces of $I_y$ versus hole density are practically identical regardless of whether $B_\parallel$ is zero or as large as $9.7$~tesla.  This clearly shows that the width of the Landau levels, and thereby the strength of the disorder potential in the system do not change with the parallel component of the magnetic field.

One of the models suggested to account for the giant magnetoresistance in metallic 2D systems was based on the Fermi surface distortion in the presence of the parallel magnetic field \cite{dassarma00}.  Our results show that parallel magnetic fields up to at least 10~tesla produce no effect on the density of states, and therefore do not noticeably affect the Fermi surface, in a p-GaAs heterostructure grown on (100) surfaces.

\begin{figure}\vspace{-28.5mm}
\begin{center}
\scalebox{0.38}{\includegraphics{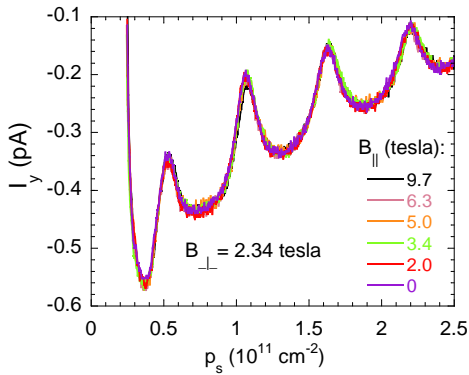}}\vspace{-2.5cm}
\caption{\label{fig.4.} $I_y$ {\it vs.}\ the hole density for different tilted magnetic fields. There is no apparent dependence of $I_y$ on the parallel component of magnetic field. $T=0.2$~K, $V_{ac}=20$~mV, $f=5$~Hz.}
\end{center}
\end{figure}

As has already been mentioned, the absence of the effect of $B_\parallel$ on either density of states or conductivity suggests that the spins of the mobile carriers in our 2D system cannot be aligned by the {\em parallel} magnetic field (in contrast to the {\em perpendicular} one).  Thus, the $g$-factor in parallel field must be nearly zero, in agreement with Ref.\cite{winkler00}.  In order to estimate the parallel-field $g$-factor, we measure the spin splitting at Landau level fillings $\nu=1$ and 3 by integrating $d\mu/dp_s$ over $p_s$ in constant $B_\perp=2.34$~tesla, but in different $B_\parallel$.  To get rid of a monotonic $\mu(p_s)$ dependence which masks the gaps in the spectrum we subtract the zero-field $\mu(p_s)$ dependence (the lower solid curve in Fig.~2) from $\mu(p_s)$ obtained by integrating the curves shown in Fig.~4.  Results for $\mu$ versus $p_s$ for $B_\parallel=9.7$~tesla and $B_\parallel=0$ are shown in Fig.~5.  There is no systematic dependence of either $\nu=1$ or $\nu=3$ spin gaps on the parallel component of the field.  This confirms that the in-plane $g$-factor is close to zero.  As seen in the figure, the uncertainty in the values of the spin gaps does not exceed 5~$\mu$eV. After normalizing by 10~tesla this uncertainty corresponds to an upper bound for the $g$-factor of $~10^{-2}$.  This result supports the notion that the giant magnetoresistance observed in other strongly-interacting 2D systems (silicon metal-oxide-semiconductor field-effect transistors and p-GaAs heterostructures on (311) surface) is a spin effect: when spins cannot be aligned, no magnetoresistance is observed.  In this sense, p-GaAs heterostructures on (100) surface are similar to p-SiGe heterostructures in which spins of free carriers cannot be affected by moderate $B_\parallel$ \cite{senz99,coleridge02}.

\begin{figure}\vspace{-8mm}
\begin{center}
\scalebox{0.4}{\includegraphics{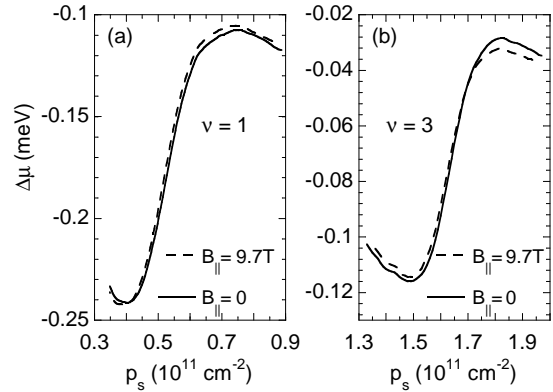}}\vspace{-4.3cm}
\end{center}
\caption{\label{fig.5.}  $\Delta\mu(p_s)$ around $\nu=$~1~(a) and 3~(b) in a constant $B_\perp=2.34$~tesla.  Solid and dashed curves correspond to $B_\parallel=0$ and $B_\parallel=9.7$~tesla, correspondingly.  $T=0.2$~K.  The curves are practically insensitive to the parallel component of the field pointing to the fact that the effective $g$-factor is close to zero in the magnetic field parallel to the surface.}
\end{figure}

In conclusion, we have shown that the compressibility (or the thermodynamic density of states) in a 2D system in p-GaAs/AlGaAs heterostructure grown on a (100) surface is insensitive to the application of a parallel magnetic field as large as 10~tesla.  The fact that the width of the Landau levels does not depend on $B_\parallel$ suggests that the disorder potential is also not sensitive to the parallel field.  In agreement with recent calculations \cite{winkler00} we conclude that the $g$-factor in this 2D system is nearly zero in the parallel magnetic field and our estimates give $g\lesssim10^{-2}$.

We are grateful to V.~T. Dolgopolov, D. Heiman, and A.~A. Shashkin for useful discussions.  This work was supported by NSF grants DMR-9988283 (Northeastern) and DMR-0071969 (UCLA) and the Sloan Foundation.

\end{document}